# Iron Embedded Magnetic Nanodiamonds for *in vivo* MRI Contrast Enhancement

Bo-Rong Lin[1,2], Chien-Hsu Chen[2], Chun-Hsiang Chang[3,4], Srinivasu Kunuku[2], Tzung-Yuang Chen[5], Tung-Yuan Hsiao[2], Hung-Kai Yu[2], Yu-Jen Chang[6], Li-Chuan Liao[6], Fang-Hsin Chen[4,7,8], Huan Niu[2*] and Chien-Ping Lee[1]

[1]Department of Electronics Engineering and Institute of Electronics, National Chiao Tung University, Hsinchu, Taiwan
[2]Accelerator Laboratory, Nuclear Science and Technology Development Center, National Tsing Hua University, Hsinchu, Taiwan
[3]Department of Biomedical Engineering and Environmental Sciences, National Tsing Hua University, Hsinchu, Taiwan
[4]Radiation Biology Research Center, Institute for Radiological Research, Chang Gung University/Chang Gung Memorial Hospital, Taoyuan, Taiwan
[5]Health Physics Division, Nuclear Science and Technology Development Center, National Tsing Hua University, Hsinchu, Taiwan
[6]Bioresource Collection and Research Center, Food Industry Research and Development Institute, Hsinchu, Taiwan
[7]Department of Radiation Oncology, Chang Gung Memorial Hospital, Taoyuan, Taiwan
[8]Department of Medical Imaging and Radiological Sciences, Chang Gung University, Taoyuan, Taiwan

Although nanodiamonds have long being considered as a potential tool for biomedical research, the practical *in vivo* application of nanodiamonds remains relatively unexplored. In this paper, we present the first application of *in vivo* MRI contrast enhancement using only iron embedded magnetic nanodiamonds. MR image enhancement was clearly demonstrated in the rendering of T2-weighted images of mice obtained using an unmodified commercial MRI scanner. The excellent contrast obtained using these nanodiamonds opens the door to the non-invasive *in vivo* tracking of NDs and image enhancement. In the future, one can apply these magnetic nanodiamonds together with surface modifications to facilitate drug delivery, targeted therapy, localized thermal treatment, and diagnostic imaging.

*Introduction*

Nanodiamonds (NDs) are highly valued for use in a variety of fields, such as lubricant additives[1], electronics devices for energy harvesting[2], slow-neutron reflectors[3], and biomedical applications[4,5]. The outstanding bio-compatibility of NDs make them ideal for drug delivery[6], anticancer therapy[7], tooth and bone implants[8], antimicrobial agents[9] and labeling[10]. ND labeling applications can be divided into two main categories: light-based methods (e.g., fluorescence imaging[11], near-infrared imaging[12], and photoacoustic imaging[13]) and magnetic resonance imaging[14-17] (MRI). When using light-based methods, defects within the diamond facilitate the emission of light and in so doing overcome many of the difficulties involved in detection. More than one hundred types of diamond defect have been identified[18], covering the entire visible spectrum. The most well-known defects are the silicon vacancy center and the nitrogen vacancy (NV) center, which produce emissions in the red region (600 ~ 800 nm), making them highly visible in fluorescence microscopy[19]. Unlike chemical dyes, the fluorescence of diamonds does not diminish over time, which makes them ideal for experiments of extended duration. In a previous paper, we presented iron embedded magnetic NDs associated with NV centers, which are compatible with existing fluorescence observation technology as a potential dual functional *in vivo* tracer[20]. Nonetheless, the *in vivo* imaging of NDs is still limited to regions that are optically accessible, i.e., only few centimeters from the surface. Imaging deep within the body requires MRI.

The ability to differentiate between normal and tumor tissue makes MRI a powerful diagnostic tool. Furthermore, MRI is noninvasive. Nonetheless, obtaining reliable images depends on the use of contrast agents. Two ND-based methods have been developed to enhance the contrast of MRI images. The first method involves attaching magnetic elements[17,21] (Gd or Mn complex) to the surface of NDs. The second method involves hyperpolarizing the NDs via the Overhauser effect[16]. Unfortunately, both of these methods necessitate complex chemical synthesis processes or specially designed ultra-low magnetic field MRI scanners. At present, there is a notable lack of research on *in vivo* MRI contrast enhancement methods using only NDs. In fact, the *in vivo* use of NDs is rare.

Ion implantation is a mature technology, which has been used in the semiconductor industry for more than 50 years[22]. This method involves the insertion of atoms directly into the diamond lattice to enhance their functionality. In a previous paper, we sought to eliminate the need for conventional chemical methods by fabricating iron embedded magnetic NDs with great bio-compatibility via ion implantation[15,23]. In this paper, we present iron embedded magnetic NDs for use as an MRI contrast agent at room temperature with an existing commercial MRI scanner. *In vivo* experiments demonstrate that the proposed iron embedded magnetic NDs are indeed nontoxic and useful, thereby opening the door on the noninvasive tracking of NDs and the immense possibilities that this brings.

## *Results and Discussions*

In a previous paper, we illustrated the efficacy of iron embedded magnetic NDs (Fe-NDs) for the *in vitro* enhancement of contrast in MRI images[15]. More advanced *in vivo* experiments require the dispersion of Fe-NDs in physiological solutions for injection into animals. Unfortunately, nanoparticles are highly susceptible to aggregation in physiological solutions[24], which greatly increases the risk of side-effects. The use of albumin to facilitate the dispersion of nanoparticles in physiological solutions is well-established[25]. Bovine serum albumin (BSA) has proven innocuous for mice[25]; therefore, we adopted BSA for the surface treatment and produced four samples: pure NDs, Fe-NDs, NDs/BSA conjugates, and Fe-NDs/BSA conjugates.

Fourier transform infrared spectroscopy (FTIR) was used to verify the attachment of the BSA. Figure 1 presents the FTIR absorbance spectra, as follows: (a) pure BSA (b) NDs/BSA conjugates, and (c) Fe-NDs/BSA conjugates. The amide characteristic peaks[26] (amide I and amide II) clearly indicate the attachment of BSA on the surface. Dynamic light scattering (DLS) was used to measure the particle size distribution and zeta potential. Figure 2 (a) presents the particle size distribution of four samples with concentration 1 *mg/ml*. Without the BSA coating, pure NDs and Fe-NDs aggregated strongly in phosphate buffered saline (PBS) at the micrometer level. The attachment of BSA significantly reduced the average particle size in the PBS to less than 400 nm. Figure 2 (b) presents the corresponding zeta potential. The attachment of BSA increased the zeta potential, due to the electrically positive nature of BSA. This provided additional evidence of BSA surface attachment. Overall, these findings clearly demonstrate the efficacy of a BSA coating in enhancing dispersibility in the solution and preventing the aggregation.

To obtain initial verification of the *in vivo* MRI contrast enhancement capability of Fe-NDs, 7T T2-weighted MR images were captured after performing simple intramuscular injection of the leg. Figure 3 presents axial and coronal T2-weighted images of C57BL/6 mice after injecting Fe-NDs (red arrow, left leg) and NDs (yellow arrow, right leg). The injection solutions containing the Fe-NDs or NDs caused the leg to swell with water. In T2-weighted images, higher water content results in images of greater brightness, which means that even if contrast enhancement does not occur, signal intensity from the leg area will become brighter after injection. We observed an obvious reduction in signal intensity in the area injected with Fe-NDs, indicating negative contrast enhancement. The untreated NDs did not produce obvious negative contrast enhancement. The success of contrast enhancement and the fact that the mice remained healthy and strong after the experiments prompted us to perform more advanced *in vivo* tests.

Previous research exploring the biodistribution of NDs revealed that the injected NDs collect in the liver and remain there for a short period of time[25,27]. In our previous FTIR tests, there was no significant difference between the Fe-NDs and NDs in terms of surface profile, which suggests that the biodistribution is probably the same. We expected that the magnetic Fe-NDs would accumulate in the liver causing local inhomogeneities in the magnetic field, and this was the case in the experiments. Water protons in the liver were dephased by inhomogeneity in the magnetic field and the T2 was shortened by additional spin-spin interactions. This led to a decrease in the signal intensity from the liver, resulting in a darker image, thereby physically confirming the contrast enhancement capability of Fe-NDs.

In this work, intraperitoneal injection was used to observe negative contrast enhancement in the mice liver because it was very difficult for us to perform intravenous injection on mice. We initially intended to follow a protocol described in the literature[27] in which the dose required for visualization is achieved through multiple intraperitoneal injections. Unfortunately, we were unable to secure the use of the MRI to perform scanning after each injection; therefore, we had to complete the *in vivo* experiments using a single high-dose injection. In the first round of experiments, we obtained T2-weighted mice MR images prior to injection, and then at one day and five days after injection. Two sets of samples were prepared. One contained only NDs/BSA conjugates. The other was Fe-NDs/BSA conjugates. Both sets were prepared at a concentration of 10 *mg/ml*. After sample injection, mouse behavior remained normal. Figure 4 presents T2-weighted C57BL/6 axial images of mice before treatment and at one day and five days after the injection of NDs or Fe-NDs. The Fe-NDs did not produce a significant increase in contrast in the original gray-level images (Figure 4 (a)). After simple imaging processing (Figure 4 (b)), the livers of the mice presented a slight decrease in signal intensity (blue arrow) one day after Fe-NDs injection. No contrast enhancement was observed at five days after Fe-NDs injection, perhaps due to the elimination of Fe-NDs from the liver. This suggests that contrast enhancement can only be achieved between one and five days after injection. Thus, in the second round of experiments, all T2-weighted MR images were obtained prior to injection and at three days after injection. Three sets of samples were prepared for the second round of experiments. One set contained only PBS as a reference, the second set contained NDs/BSA conjugates (10 *mg/ml*), and the third set included Fe-NDs/BSA conjugates (10 *mg/ml*). After sample injection, mouse behavior remained normal. Figure 5 presents T2-weighted C57BL/6 images obtained before treatment and at three days after sample injection. The only sample that presented an obvious negative contrast enhancement was the one involving Fe-NDs injection (Figure 5 (c)). In that image, the liver could be distinguished easily and the boundary between the liver and muscle was very clear (red arrow). The accumulation of Fe-NDs in the liver created local magnetic field inhomogeneities, shortened the T2 value, and enhanced the negative contrast. Any contrast enhancement provided by regular NDs and PBS (Figure 5 (a) and (b)) was insignificant.

To confirm the accumulation of NDs and Fe-NDs in the liver, further experiments were conducted on three new mice. This involved the intraperitoneal injection of PBS, NDs and Fe-NDs at the concentrations mentioned previously, whereupon the mice were sacrificed at three days after injection to obtain liver tissue for observation under microscope. Figure 6 presents the liver pathology in mice at three days after intraperitoneal administration. No nanoparticles were detected in livers injected with PBS (Figure 6 (a)). Large enough nanoparticle accumulations were clearly detected with red fluorescence in livers injected with NDs and Fe-NDs (Figure 6 (b) and (c)). The existence of NDs in the liver had little effect on MRI signal intensity (Figure 5 (b)), due to the non-magnetic nature of NDs. The accumulation of Fe-NDs in the liver (as observed in Figure 6 (c)) indicates that the change in signal intensity was due to Fe-NDs injection. These findings verify the *in vivo* enhancement of contrast by Fe-NDs from a physical perspective.

*Conclusion*

This paper demonstrates the *in vivo* enhancement of contrast through the injection of iron embedded magnetic NDs in images obtained using a typical commercial MRI scanner. This is the first research work on *in vivo* MRI contrast enhancement induced entirely by NDs. iron embedded magnetic NDs have considerable potential in imaging, targeted cancer therapy, and localized treatment all at the same time. Eventually, we hope that this new material, iron embedded magnetic NDs, will create or pave the way to cutting-edge new research topics which will help consolidate the study of prosperous NDs.

*Methods*

Fabrication of Iron embedded magnetic nanodiamonds.

ND powder (average 100nm in diameter) from Microdiamant Co. was first dissolved in DI water. The solution was then applied onto an oxidized silicon wafer and dried. The wafer was iron-ion implanted with an energy 150 *keV* and dose $5 \times 10^{15}$ *atoms/cm²*. The methods to remove the implanted NDs from the silicon wafer and to collect the iron embedded magnetic NDs have been described in our previous paper[23].

Preparation of Fe-NDs/BSA conjugates and NDs/BSA conjugates.

Both Fe-NDs and NDs were purified and carboxylated to remove the surface graphite fractions and impurities with similar previously described methods[28]. The as received NDs and Fe-NDs were treated with a mixture solution of $H_2SO_4$ and $HNO_3$ (volume ratio 9:1) at room temperature for 24 hours respectively. After reactions, the solutions were centrifuged to get purified Fe-NDs and NDs. Purified Fe-NDs and NDs were added into the 0.1 M NaOH solution at 90 °C for 2 hours and next 0.1 M HCl solution at 90 °C for 2 hours respectively. After reactions, the solutions were centrifuged to get carboxylated Fe-NDs and NDs. Both of them were washed many times with DI water to remove free acid. After drying process, 10 *mg* of carboxylated Fe-NDs and NDs were mixed together with 50 *mg* BSA powder from Sigma-Aldrich in 2 *ml* PBS respectively. After thorough mixing, the solutions were centrifuged and the sediment was washed several times to get Fe-NDs/BSA conjugates and NDs/BSA conjugates.

FTIR measurement.

To confirm the attachment of BSA, three samples, pure BSA, Fe-NDs/BSA conjugates and NDs/BSA conjugates were coated onto the silicon wafers and the absorbance of each samples were measured using a fourier transform infrared spectrometer (Bruker IFS66V/S). It is worth noting that silicon has no absorption within the measured range.

Particle size distribution and Zeta potential measurement

To understand the effect of the BSA attachment, four samples, Fe-NDs, NDs, Fe-NDs/BSA conjugates and NDs/BSA conjugates, were dispersed in PBS with concentration 1 *mg/ml*. The particle size distribution and Zeta potential of all samples were measured using a Zetasizer Nano ZS from Malvern Instruments.

*In vivo* MR imaging through intramuscular injection

All animal experimental procedures and studies were approved by and in accordance with the Institutional Animal Care and Use Committee (IACUC) of National Tsing Hua University. The IACUC protocol number was 107028. The C57BL/6 mice (8-week-old male) were obtained from the Laboratory Animal Center of NAR Labs (NAR Labs, Taipei, Taiwan). Fe-NDs were dispersed in PBS containing 30% Matrigel from CORNING with concentration 2.4 *mg/ml*. Control sample was regular NDs with the same concentrations. Bruker BIOSPEC 70/30 MRI scanner equipped with proper gradient coils was used. RARE pulse sequence with TR = 2700 *ms*, TE = 26.7 *ms*, matrix size = 256 x 256, field of view = 40 x 40 $mm^2$, slice thickness = 1 *mm*, an average of 4 was used to capture T2-weighted images. Fe-NDs and NDs samples were intramuscular injected into the left and right leg of mice respectively. T2-weighted images were performed from coronal and axial direction to see the contrast enhancement.

*In vivo* MR imaging through intraperitoneal injection

Fe-NDs/BSA conjugates was dispersed in PBS with concentration 10 *mg/ml*. Control samples were NDs/BSA conjugates with the same concentration and PBS. Bruker BIOSPEC 70/30 MRI scanner equipped with proper gradient coils was used. RARE pulse sequence with TR = 3100 *ms*, TE = 30 *ms*, matrix size = 256 x 256, field of view = 40 x 40 $mm^2$, slice thickness = 0.5 *mm*, an average of 8 was used to capture liver T2-weighted images. Following the pre-scan, Fe-NDs/BSA conjugates, control NDs/BSA conjugates and PBS samples 1 *ml* were intraperitoneal injected into the mice respectively. After absorption of planed days, liver T2-weighted images were performed again to see the contrast enhancement.

Pathology and biodistribution

NDs/BSA and Fe-NDs/BSA conjugates were dispersed in PBS buffer to a final concentration of 10 *mg/ml* and a volume of 1 *ml* was intraperitoneally injected into mice. After 72 hours, mice were sacrificed and the liver tissues were embedded in optimal cutting temperature compound and then stored in -80 °C. The liver tissues were cryogenic-sectioned from transverse axis for 10 *um* in thickness. After fixation in -20 °C methanol for 5 minutes, the tissues were washed twice and then mounted by DAPI reagent (4',6-diamidino-2-phenylindole, Invitrogen) to visualize the nuclei. The biodistribution of the injected NDs and Fe-NDs were detected by its autofluorescence and visualized via microscope equipped with 605±55 *nm* filter.

Figure captions

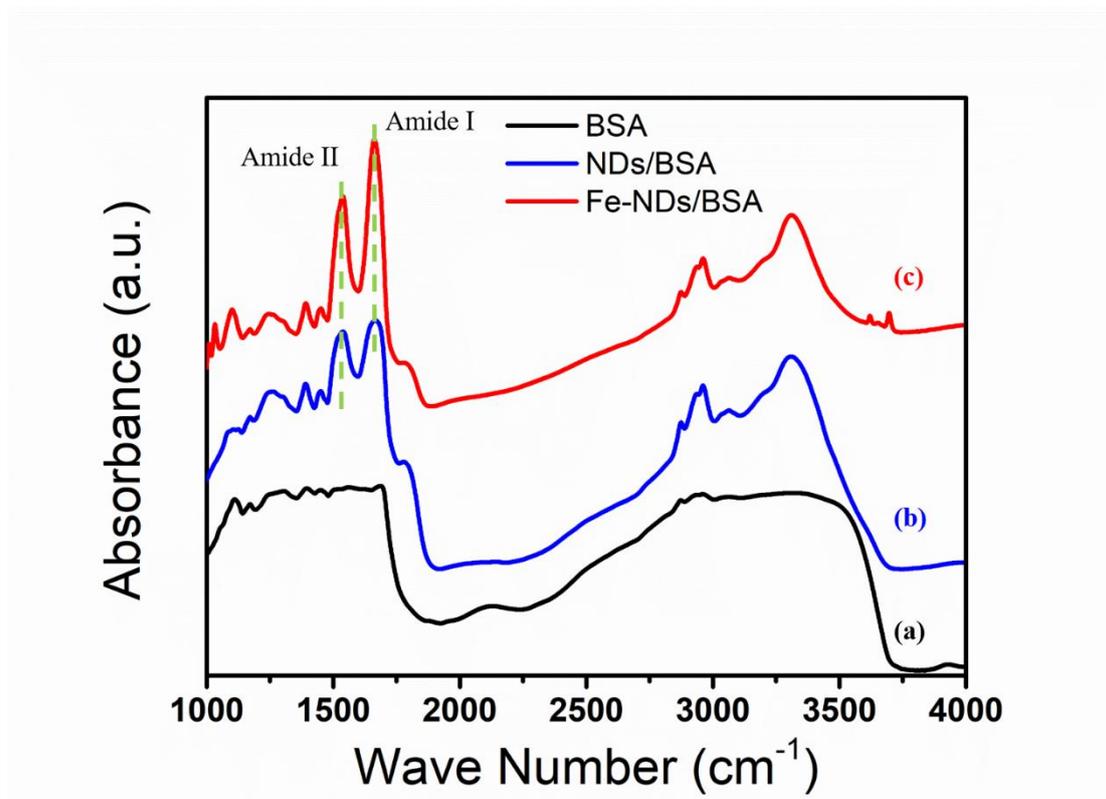

Figure 1. Fourier Transform Infrared (FTIR) spectra (a) BSA, (b) NDs/BSA conjugates and (c) Fe-NDs/BSA conjugates. The amide characteristic peaks (amide I and amide II) clearly indicate the attachment of BSA on the surface of Fe-NDs and NDs.

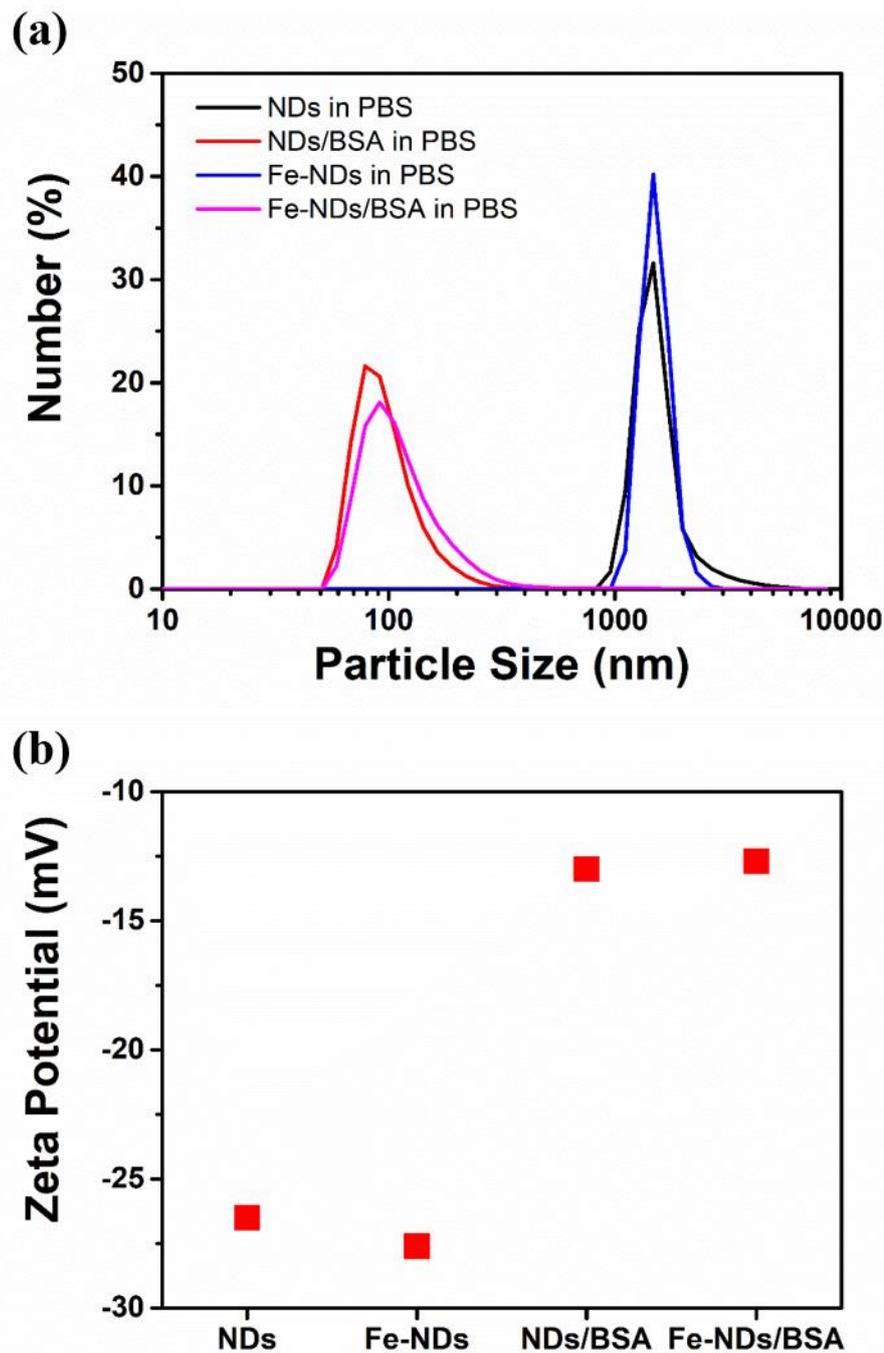

Figure 2. (a) Dynamic light scattering particle size measurements of NDs, Fe-NDs, NDs/BSA conjugates and Fe-NDs/BAS conjugates in PBS. NDs and Fe-NDs dispersed in PBS have a large aggregation size larger than 1000 *nm*. After BSA adsorption on the NDs/Fe-NDs surface, both NDs/BSA conjugates and Fe-NDs/BAS conjugates are better dispersed in PBS. (b) Zeta potential measurements of NDs, Fe-NDs, NDs/BSA conjugates and Fe-NDs/BAS conjugates in PBS. After BSA attachment, the zeta potentials increased due to the electric positive nature of BSA.

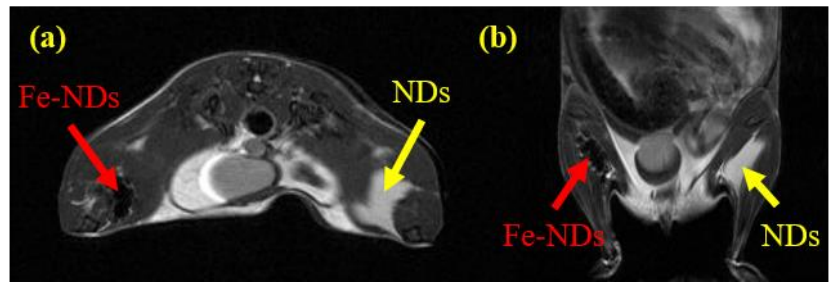

Figure 3. (a) Axial and (b) coronal T2-weighted image with TR = 2700 *ms* and TE = 26.7 *ms* after Fe-NDs (red, left leg) and NDs (yellow, right leg) intramuscular injection. Clear contrast enhancement was clearly observed at Fe-NDs injected left leg.

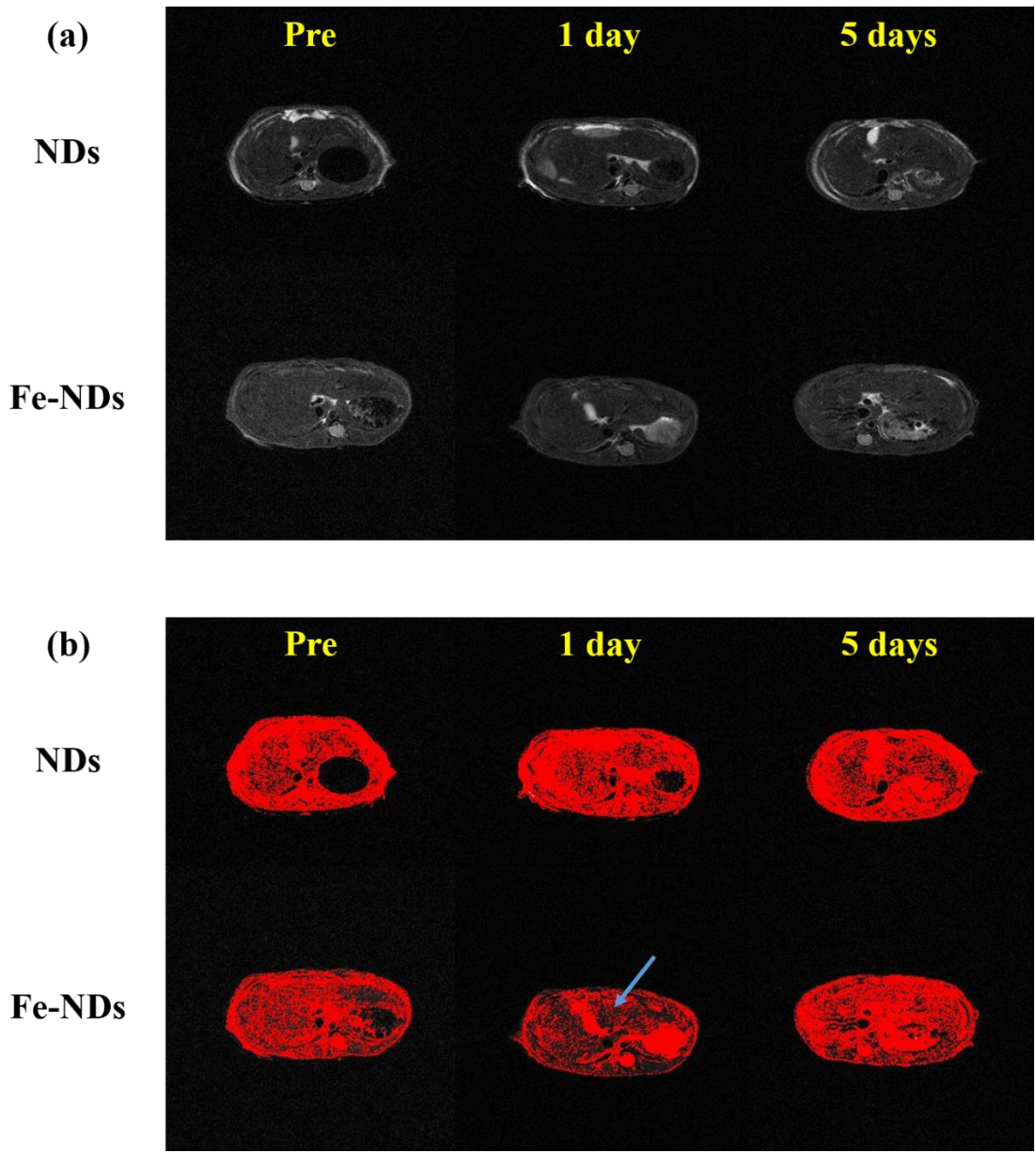

Figure 4. (a) Axial liver T2-weighted images with TR = 3100 *ms* and TE = 30 *ms* without imaging processing at different time after NDs and Fe-NDs intraperitoneal injection. (b) Corresponding axial liver T2-weighted images with imaging processing. Images are shown pre, a day and 5 days after injection. Negative contrast enhancement was observed at Fe-NDs injected mouse after imaging processing.

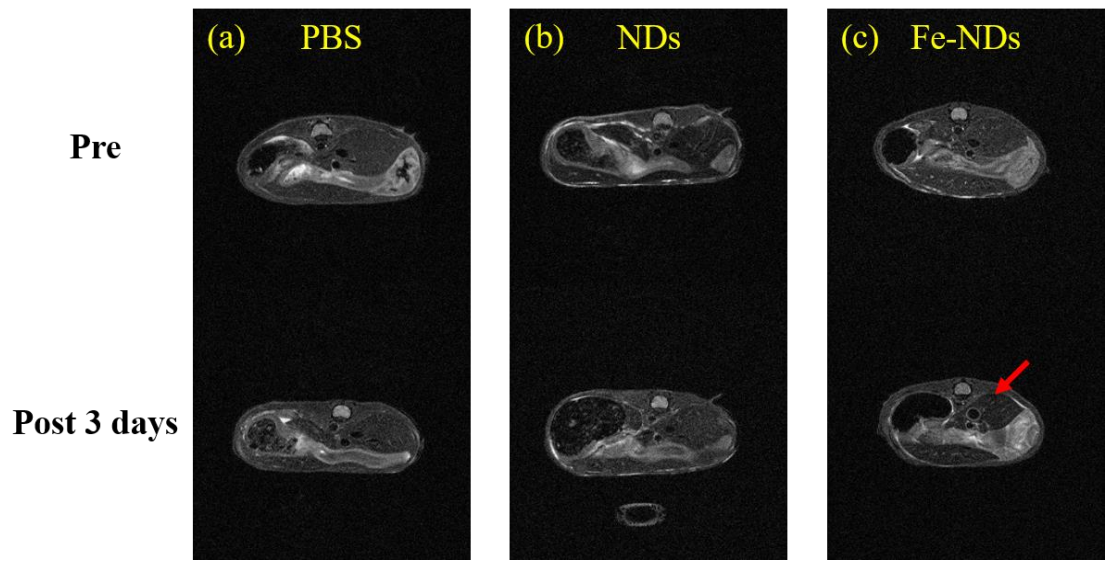

Figure 5. Axial liver T2-weighted images with TR = 3100 *ms* and TE = 30 *ms* after (a) PBS, (b) NDs and (c) Fe-NDs 3 days intraperitoneal injection. Clear negative contrast enhancement was clearly observed at Fe-NDs injected mouse.

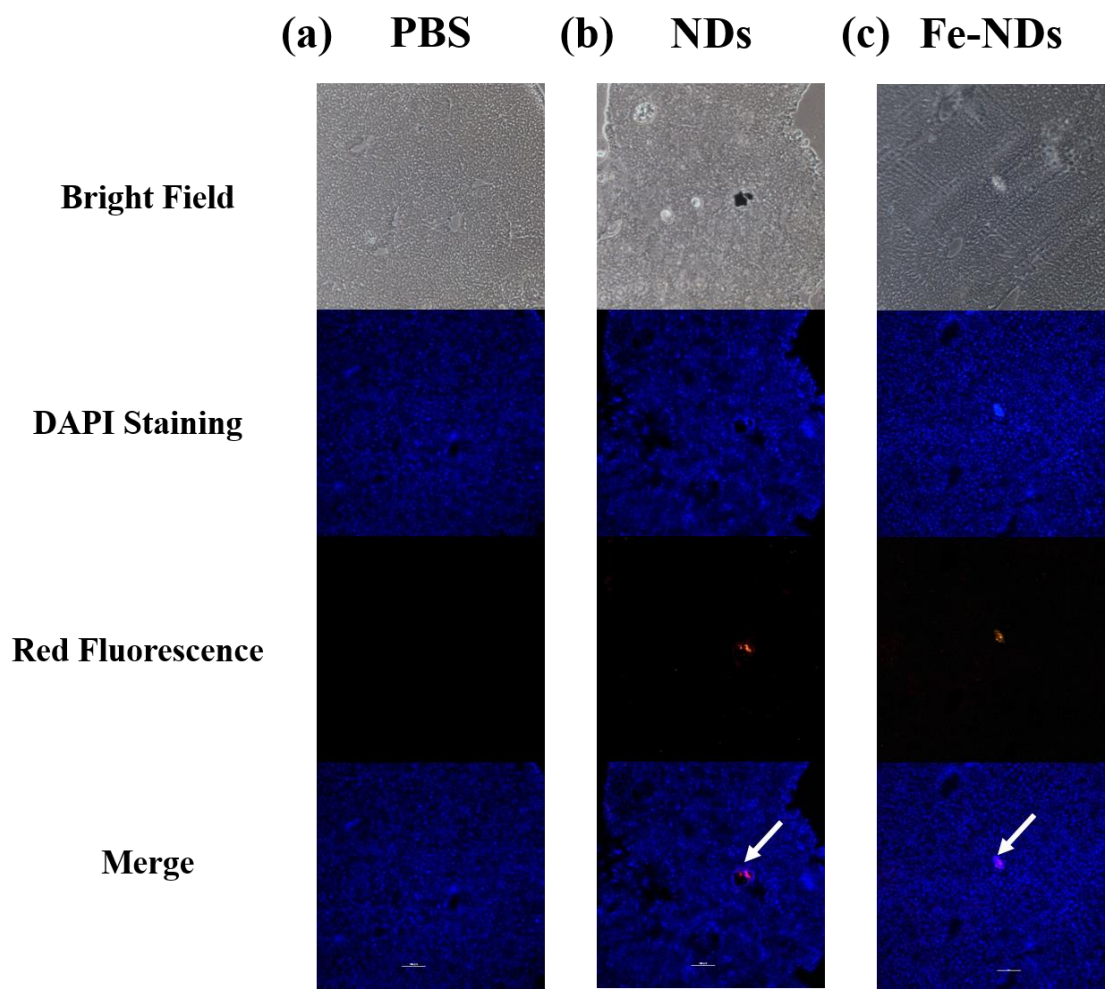

Figure 6. Mice liver bright field pathology, corresponding DAPI staining and red fluorescence images after (a) PBS, (b) NDs and (c) Fe-NDs 3 days intraperitoneal injection. Nanoparticle accumulations were clearly observed in livers injected with NDs and Fe-NDs. Scale bar: 100 $\mu m$.

Acknowledgements

This work was financially supported by Team Union Ltd.